\documentclass[prb,amsmath,amssymb,twocolumn,showpacs,floatfix]{revtex4}
\usepackage{graphicx}
\usepackage[english]{babel}
\usepackage{times}
\usepackage{dcolumn}
\usepackage[usenames,dvipsnames]{color}
\usepackage{units}

\begin{document}

\title{Magnetic anisotropy effects on quantum impurities in
superconducting host}

\author{Rok \v{Z}itko $^{1,2}$, Oliver Bodensiek $^3$, Thomas
Pruschke $^3$}

\affiliation{$^1$ Jo\v{z}ef Stefan Institute, Jamova 39, SI-1000 Ljubljana,
Slovenia,\\
$^2$ Faculty  of Mathematics and Physics, University of Ljubljana,
Jadranska 19, SI-1000 Ljubljana, Slovenia,\\
$^3$ Institute for Theoretical Physics, University of G\"ottingen,
Friedrich-Hund-Platz 1, D-37077 G\"ottingen, Germany}

\date{\today}

\begin{abstract}
We study the magnetic anisotropy effects on the localized sub-gap
excitations induced by quantum impurities coupled to a superconducting
host. We establish the ground-state phase diagrams for single-channel
and two-channel high-spin Kondo impurities; they unveil surprising
complexity that results from the (multi-stage) Kondo screening in
competition with the superconducting correlations and the magnetic
anisotropy splitting of the spin multiplets. We discuss the
possibility of detecting the Zeeman splitting of the sub-gap states,
which would provide an interesting spectroscopic tool for studying the
magnetism on the single-atom level. We also study the problem of two
impurities coupled by the Heisenberg exchange interaction, and we
follow the evolution of the sub-gap states for both antiferromagnetic
and ferromagnetic coupling. For sufficiently strong antiferromagnetic
coupling, the impurities bind into a singlet state that is
non-magnetic, thus the sub-gap states move to the edge of the gap and
can no longer be discerned. For ferromagnetic coupling, some excited
states remain present inside the gap. 
\end{abstract}

\pacs{72.10.Fk, 72.15.Qm, 73.20.Hb, 73.20.-r}

\maketitle

\newcommand{\vc}[1]{{\mathbf{#1}}}
\newcommand{\vck}{\vc{k}}
\newcommand{\braket}[2]{\langle#1|#2\rangle}
\newcommand{\expv}[1]{\langle #1 \rangle}
\newcommand{\ket}[1]{| #1 \rangle}

\section{Introduction}

\begin{figure}[htbp]\centering
\includegraphics[clip,width=6cm]{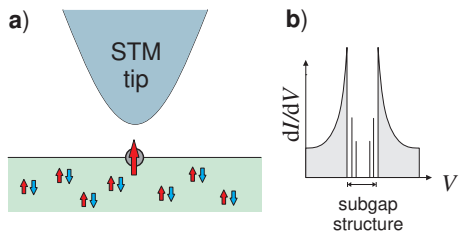}
\caption{(Color online)
(a) Magnetic impurity atom adsorbed on the superconductor
surface probed by the STM.
(b) Idealization of the characteristic differential conductance spectrum
as recorded by the STM above a magnetic impurity: in the gap,
the LDOS is essentially zero except for the discrete peaks which
correspond to the transitions from the ground state 
to the sub-gap excited states.
}
\label{stm}
\end{figure}

Tunneling spectroscopy is the prevalent experimental approach for
studying superconductivity. It provides information on such
fundamental properties as the energy gap, pairing symmetry, and
pairing interactions \cite{giaever1960, giaever1974, pan1999zn,
yazdani1999, balatsky2006, fischer2007}. Using a scanning tunneling
microscope (STM) it is possible to examine the impurity effects on the
single-atom level \cite{binnig1982, lang1986, crommie1993c,
yazdani1997, yazdani1999, derro2002, heinrich2004, stroscio2004,
neel2007, yayon2007, alloul2009, brune2009, ternes2009}; see
Fig.~\ref{stm}a). Such measurements provide crucial data on the nature
of the superconducting state in complex materials \cite{pan1999zn,
balatsky2006}. With improvements in the instrumentation, experiments
are being performed at increasingly low temperatures and ever further
details in the local density of states (LDOS) can be resolved: recent
STM work performed in the $\unit[300]{mK}$ range on magnetic adatoms
adsorbed on superconductors has clearly revealed the existence of
multiple sub-gap excitation peaks \cite{ji2008, iavarone2010, ji2010};
see Fig.~\ref{stm}b). It was proposed that these may be interpreted as
the magnetic-impurity-induced bound states associated with the
different angular-momentum scattering channels \cite{ji2008,
flatte1997prl, flatte1997prb}, but we show in this work that an
alternative interpretation in terms of the magnetic-anisotropy effects
is also possible.

Since the magnitude of the magnetic anisotropy of magnetic adsorbates
is comparable or may even exceed the superconducting gap
\cite{gambardella2003, hirjibehedin2007}, its effects are drastic, yet
they have received hardly any attention in this context. In addition, it
has been very recently demonstrated that the strength of the
anisotropy (parameter $D$) in magnetic molecules can be continuously
tuned in mechanical break junctions \cite{parks2010}; using
superconducting leads, one could then directly study the effect of the
magnetic anisotropy on the transport and the spectral properties of
magnetic molecules coupled to superconductors. Finally, spin-orbit
coupling is always present for any impurity atom embedded in the
superconducting bulk, therefore the class of problems in which an
anisotropic spin degree of freedom is coupled to a superconducting
host is indeed wide ranging.

The theory of magnetic impurities in a superconductor was first worked
out within simplified classical-spin models \cite{shiba1968,
sakurai1970}, while later works took into account the quantum nature
of impurities and, among other improvements, properly described the
competition between the screening of the impurity moment by the Kondo
effect and the superconducting correlations \cite{zmha}. The crucial
difference between a classical and quantum spin is that a classical
spin is static (essentially equivalent to a local point-like magnetic
field), it has no internal dynamics, and it cannot flip the spin of
the conduction-band electrons. For this reason, the impurity problem
in a classical-spin approximation is a non-interacting (mean-field)
theory. The quantum impurity, however, needs to be treated using the
tools of the many-particle theory, which can take into account the
non-perturbative effects induced by strong interactions. The behavior
of classical and quantum impurities is very different in many
respects. For classical impurities, there is no difference if the
coupling to the conduction-band electrons is ferromagnetic or
antiferromagnetic, while real quantum impurities have very different
properties in these two cases. Furthermore, in the absence of magnetic
anisotropy, a quantum impurity model has full SU(2) spin symmetry and
any breaking of this rotational invariance would indicate a deficiency
of the method used; a classical spin, however, explicitly breaks the
spin symmetry even at the level of the Hamiltonian itself. Unless
there is a physical mechanism that can lead to a real symmetry
breaking [such as spontaneous symmetry breaking due to magnetic
ordering induced by inter-impurity Ruderman-Kittel-Kasuya-Yosida
(RKKY) interaction], only a quantum impurity model can thus provide
the qualitatively correct result, while a classical model will be
affected by spurious symmetry breaking. (In magnetically-ordered
systems, classical spin models may be fully adequate for many
purposes.) The spin SU(2) symmetry breaking also implies the breaking
of the time-reversal symmetry. This has important consequences for the
degeneracy of the sub-gap states (cf. Kramers' degeneracy theorem) and
it is one of the major differences between quantum and classical
(static) impurities. This dichotomy recently gained renewed attention
in the context of topological insulators \cite{moore2010, fu2007,
hsieh2009}, since only impurities which break the time-reversal
invariance can open the gap in the topologically-protected
edge/surface states \cite{roushan2009, fu2009, alpichshev2010,
zhangT2009, guo2010, liu2009, zhou2009, feng2009, topo, cha2010}.

Accurate calculations for quantum impurities in superconductors became
possible by generalizing the numerical renormalization group (NRG)
method to problems with a superconducting electron bath
\cite{satori1992}. This method was applied to simplified models such
as the spin-$1/2$ Kondo model \cite{satori1992, sakai1993,
yoshioka1998} and the non-degenerate Anderson impurity model
\cite{yoshioka2000}. Real magnetic impurity atoms and molecules
generally require, however, a more sophisticated description in terms
of the multi-channel degenerate Anderson model or the high-spin Kondo
model with magnetic anisotropy terms \cite{otte2008}. The goal of the
present work is thus to apply the NRG to study the sub-gap excitations
for such more realistic models, featuring anisotropic high-spin Kondo
impurities with more than one channel. There are some related works in
the literature. Lee et al. have studied the isotropic high-spin
impurities in a side-coupled configuration \cite{lee2008} and the
two-level impurity model \cite{lee2010}. The two-channel models have
been studied in the context of unconventional superconductors
\cite{koga2002sc} and iron pnictide superconductors
\cite{matsumoto2009}. Moca et al. have demonstrated the presence of
multiple sub-gap states in a multi-orbital model for Mn impurity in
MgB$_2$ \cite{moca2008}. Multiple sub-gap states also appear in the
case of XXZ anisotropic Kondo exchange coupling in the
$S_\mathrm{imp}=1/2$ Kondo model \cite{yoshioka1998}. Multi-channel
high-spin Kondo models with magnetic anisotropy terms, however, have
not yet been studied.

The paper is structured as follows. In Sec.~\ref{sec2}, we define the
model and comment on its relevance for actual adsorbate systems.  In
Sec.~\ref{sec3}, we study the ground state diagrams for various one-
and two-channel isotropic high-spin Kondo systems, while the magnetic
anisotropy effects are presented in Sec.~\ref{sec4}. The spectral
peaks and their splitting due to the magnetic anisotropy in the
multi-channel case are discussed in Sec.~\ref{sec5}. Section~\ref{sec6}
is devoted to the role of the external magnetic field, which splits
all many-particle states with $S \neq 0$. Finally, impurity dimers
are studied in Sec.~\ref{sec7}.

\section{Model and method}
\label{sec2}

We describe the impurity system by the Hamiltonian $H=H_\mathrm{band}
+ H_\mathrm{imp}$, where $H_\mathrm{band}$ describes $N$ channels of
conduction-band electrons using the mean-field BCS Hamiltonian with
the gap $\Delta$ \cite{satori1992}:
\begin{equation*}
H_\mathrm{band} = \sum_{i=1}^N \left[ \sum_{k\sigma}
\epsilon_k c^\dag_{k\sigma i} c_{k \sigma i}
+ \sum_k \Delta \left( c_{k \uparrow i}^\dag
c_{k \downarrow i}^\dag + \text{H.c.} \right) \right],
\end{equation*}
while the impurity is described by a Kondo-like Hamiltonian with a
magnetic anisotropy term \cite{otte2008, aniso, aniso2, romeike2006,
leuenberger2006, gonzalez2008, ujsaghy1996}:
\begin{equation*}
H_{\mathrm{imp}} = \sum_{i=1}^N J_i \vc{S}_\mathrm{imp} \cdot \vc{s}_i 
+ D \vc{S}_{\mathrm{imp},z}^2 + g \mu_B B S_{\mathrm{imp},z}.
\end{equation*}
Here $J_i$ are the exchange coupling constants, $\vc{S}_\mathrm{imp}$
is the impurity spin operator satisfying the su(2) Lie algebra
$[S_{\mathrm{imp},\alpha}, S_{\mathrm{imp},\beta}]=i
\epsilon_{\alpha\beta\gamma} S_{\mathrm{imp},\gamma}$, and $\vc{s}_i$
is the channel-$i$ spin density at the position of the impurity:
\begin{equation}
\vc{s}_i = \frac{1}{\mathcal{N}} \sum_{kk'\alpha\beta} c^\dag_{k\alpha i}
\left( \frac{1}{2} \boldsymbol{\sigma}_{\alpha\beta} \right)
c_{k'\beta i},
\end{equation}
where $\mathcal{N}$ is the number of the states in the conduction
band. Furthermore, $D$ is the longitudinal anisotropy, $g$ is the
impurity $g$-factor, $\mu_B$ is the Bohr magneton, and $B$ is the external
magnetic field (non-zero $B$ is discussed in Sec.~\ref{sec6}). The
multiple channels correspond to the different symmetry channels of the
Bloch states, which hybridize with the impurity $d$-levels; our
high-spin Kondo model may be thought to arise from some multi-orbital
Anderson model after performing the Schrieffer-Wolff transformation
\cite{anderson1961, schrieffer1965, schrieffer1966, coqblin1969,
gruner1974, anderson1978, hewson}. Strictly speaking, the low-energy
effective model for a multi-orbital Anderson model is not necessarily
a high-spin Kondo model. Such an exception occurs, for example, if the
system is not in the local-moment regime, but rather its valency is
strongly fluctuating, or if there is also some orbital moment on the
impurity atom. For surface-adsorbed impurities, the symmetry in real
space is broken, thus it is reasonable to expect strong quenching of
the orbital moment. The valence-fluctuation regime cannot be excluded
a priori, but the systems of such complexity are beyond the
capabilities of the NRG method. In this work we thus focus on the
problems where the orbital moment is quenched and the electrons in the
$d$-orbitals are locked into a high-spin state by the strong Hund's
coupling. Such cases are adequately described by the proposed model.
For $D \neq 0$ or $B \neq 0$, the Hamiltonian only has an axial
$\mathrm{U}(1)$ spin symmetry, thus the sole conserved quantum number
is $S_z$, the $z$ component of the total spin.

The NRG method consists of discretizing the continua of the
conduction-band states, rewriting the Hamiltonian in the form of
one-dimensional tight-binding chains with an exponentially decreasing
hopping constants, and diagonalizing the resulting Hamiltonian
iteratively by adding one further chain site per channel in each step
\cite{wilson1975, krishna1980a, bulla2008}. The spectrum of
many-particle states is truncated to the low-energy part after each
step. Due to the low symmetry of the problem (there is no particle
conservation in the superconducting case, and there is only partial
spin symmetry in the presence of magnetic anisotropy or magnetic
field), these calculations are numerically very demanding. The size of
the matrices that need to be diagonalized at the given truncation
cutoff strongly depend on the value of the discretization parameter
$\Lambda$. In many situations, one can use a large value of $\Lambda$
and reduce the discretization artifacts by the so-called $z$-averaging
trick \cite{oliveira1994, silva1996, paula1999, campo2005,
resolution}; this approach produces excellent results for featureless
(flat) conduction bands. One needs to be very careful, however, in the
vicinity of phase transitions, since the ground state obtained in a
calculation can be $z$-dependent (in other words, it is possible that
for exactly the same model parameters, one obtains a different ground
state for different interleaved discretization meshes). The averaging
may then be ill-defined for an interval of parameters where such
$z$-dependence of the ground state occurs. The width of this interval
grows as $\Lambda$ is increased. Nevertheless, experience shows that
in spite of this difficulty, one can determine the transition point
very accurately even by performing the calculations with a very large
value of $\Lambda$. Test calculations on simple problems, for example,
show that by determining the value of the system parameter where the
transition occurs for a fixed value of $z$, and averaging such results
over $z$, one obtains a value that changes little with $\Lambda$.
Often it is sufficient to use only two values of $z$ (such as 1 and
0.5) to obtain good results. This approach has been used, for example,
to establish the accurate values tabulated in Table~\ref{tab1} in the
following section. In other parts of this work, where high accuracy
was not essential, we performed no such averaging, thus the results
are only qualitatively correct.

\section{Transitions in the isotropic model}
\label{sec3}

The properties of the multi-channel Kondo model in the normal state
depend on the relation between the number of channels and the impurity
spin; roughly speaking, each channel can screen one half unit of the
impurity spin \cite{mattis1968, cragg1979b, nozieres1980, cragg1980,
rajan1982, sacramento1989, sacramento1991, hewson}. Thus, for
$N<2S_\mathrm{imp}$, the impurity spin can be only partially screened,
while for $N=2S_\mathrm{imp}$ there is an exact spin compensation,
yielding a singlet ground state (GS); for $N>2S_\mathrm{imp}$, exotic
non-Fermi-liquid (NFL) states may arise \cite{cox1998}.  In the
superconducting state, the behavior of the Kondo model is only well
explored for the simplest case of $N=1$ and $S_\mathrm{imp}=1/2$
\cite{satori1992}: as the gap $\Delta$ is increased from zero, there
is a transition at $\Delta_c$ between the regime where the impurity is
Kondo-screened (singlet ground state) to the regime where the impurity
is free (doublet ground state). The value of $\Delta_c$ is of the
order $T_K$, the Kondo temperature, which is the characteristic energy
scale of the Kondo effect. The transition occurs because in the
superconducting state there is an insufficient number of low-energy
electron states to participate in the Kondo screening. 

We now consider the general multi-channel high-spin case, first in the
absence of the anisotropy and for equal coupling constants $J_i$ for
all channels. By analogy with the known results, we expect that for
$2S_\mathrm{imp}\geq N$ there is a transition from the Kondo screened
$S=S_\mathrm{imp}-N/2$ state to the ``free-spin'' $S=S_\mathrm{imp}$
state as the gap is increased. Our NRG calculations fully support this
picture. For $2S_\mathrm{imp} < N$ the NFL effects make a-priori
predictions difficult; in numerical simulations for $N=2$ and
$S_\mathrm{imp}=1/2$, we observe a transition from a degenerate pair of
singlet states to a doublet ground state as $\Delta$ increases. The
$\Delta_c$ for $N=1$ and $N=2$ are tabulated in Table~\ref{tab1} for a
range of $S_\mathrm{imp}$. The ratio $\Delta_c/T_K$ strongly depends
on $S_\mathrm{imp}$ in spite of the fact that the superexchange
couplings are constant and thus the Kondo scale is formally the same
in all cases. The variation with $N$ is weaker, with a notable
exception of the NFL case with $N=2$, $S_\mathrm{imp}=1/2$, where
$\Delta_c/T_K$ is much reduced.

\begin{table}[htb]
\centering
\begin{ruledtabular}
\begin{tabular}{@{}c|ccccc@{}}
$S_\mathrm{imp}$ & $1/2$ & $1$ & $3/2$ & $2$ & $5/2$ \\
\hline
$N=1$ & 3.7     & 5.9   & 10.3    & 19.8   &  43   \\
$N=2$ & 0.70    & 5.7   & 11.3    & 23     &  51 \\
\end{tabular}
\end{ruledtabular}
\caption{Dependence of the ratio $\Delta_c/T_K$ between the critical
gap and the Kondo temperature on the number of channels $N$ and the
impurity spin $S_\mathrm{imp}$. In the two-channel case, the coupling
to both channels is taken to be equal, $J_i \equiv J$. In all
calculations $J=0.2$, and the Kondo temperature is $T_K=1.16\times
10^{-5}$. We use Wilson's definition of the Kondo temperature
\cite{wilson1975} throughout this work. } \label{tab1}
\end{table}

It must be emphasized that even in the limit of small exchange
couplings $J_i$, that is, when the Kondo effect plays no role and the
impurity remains unscreened, the system still is not equivalent to a
classical spin. The ground state for small $J_i$ is a degenerate spin
$S$ multiplet, which is not equivalent to a single spin-polarized
state as predicted by a classical spin model (the symmetries of the
state are different). Another important observation is that, quite
generally, in quantum impurity models one needs $\Delta \sim T_K$ to
observe sub-gap excitations well inside the gap (i.e., not at the very
edge), as shown in Table~\ref{tab1}. Thus in the situations where the
spin-flip scattering processes play no role, we also do not expect to
observe any sub-gap peaks deep in the gap. This is in contradiction
to the results of the classical spin models, which on the one hand
presume the irrelevance of the Kondo effect due to small exchange
coupling, yet also predict excitations well inside the gap. Such
a discrepancy exists even for the relatively large spin
$S_\mathrm{imp}=5/2$, which still cannot be considered as a classical
spin.

\section{Transitions in the anisotropic case}
\label{sec4}

\begin{figure}[htbp]\centering
\includegraphics[clip,width=9cm]{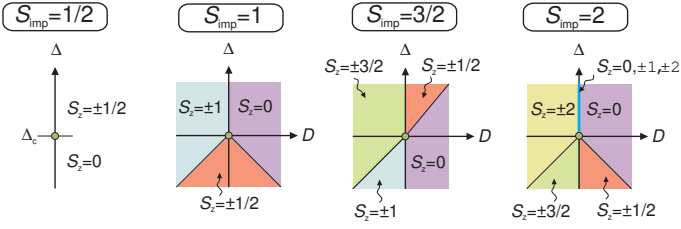}
\caption{(Color online) Many-body ground state as a function of the
gap $\Delta$ and the magnetic anisotropy $D$ at fixed exchange
coupling $J$ for a range of the impurity spin $S_\mathrm{imp}$ in the
single-channel case. Magnetic field is zero, $B=0$. The value
$\Delta_c$ (circle) corresponds to the transition point between the
Kondo screened and unscreened impurity moment in the isotropic case. 
}
\label{phase1}
\end{figure}

In the anisotropic case, the ground-state multiplet with $S \geq 1$
splits: for axial $D<0$ anisotropy the new ground state consists of
states with the maximal $S_z=\pm S$, while for planar $D>0$ anisotropy
the ground state is $S_z=0$ for integer $S$ and $S_z=\pm1/2$ for
half-integer $S$. The transition point at $\Delta_c$ in the isotropic
models is extended into transition lines in the $(D,\Delta)$ plane.
For a given value of $D$, there is some gap $\Delta$ where the system
makes a transition from a ``low-$|S_z|$'' to a ``high-$|S_z|$'' regime,
which are the equivalents of the Kondo-screened and free-spin regime,
respectively. The results of extensive calculations for the
single-channel problems are summarized in Fig.~\ref{phase1} in the
form of phase diagrams, while the actual numerical results (including
also all sub-gap excited states) are shown in the Appendix.

\begin{figure}[htbp]\centering
\includegraphics[clip,width=8.5cm]{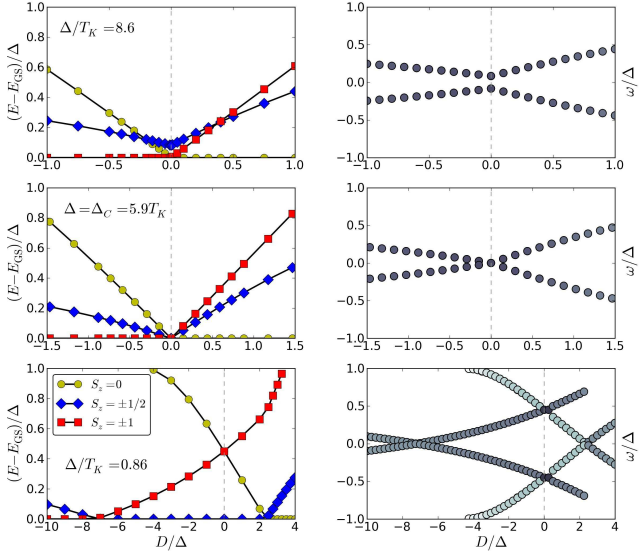}
\caption{(Color online) Left panels: Ground state and sub-gap
many-body excited states as a function of the magnetic anisotropy $D$
for different values of the superconducting gap $\Delta$ for the
single-channel spin-1 Kondo models. Zero magnetic field, $B=0$.
Top to bottom: ``free-spin''
regime, transition regime, Kondo screened regime.
The energies $E$ are plotted relative to the
ground-state energy $E_\mathrm{GS}$ and they are rescaled in units of
the gap, thus the ground state always lies atop the horizontal axis at
$E=E_\mathrm{GS}$ and the continuum of excitations starts at
$(E-E_\mathrm{GS})/\Delta=1$.\quad
Right panels: Energies and weights of the sub-gap spectral peaks in
the impurity spectral function. Darker shade corresponds to higher
spectral weight of the delta peak.}
\label{fig2}
\end{figure}

Let us consider the results for $S_\mathrm{imp}=1$ more closely. In
the left panels in Fig.~\ref{fig2} we plot the ground state as well as
the sub-gap excited-state energies as a function of the anisotropy
parameter $D$ for three values of $\Delta$ that correspond in the
isotropic limit to the Kondo regime, the transition regime, and the
``free-spin'' regime, respectively. The spin-1 ground state and
excited state (ES) spin multiplets split in the presence of the
magnetic anisotropy. The degree of splitting is not the same as for
free multiplets, but rather depends on $\Delta$ and $T_K$. For
example, at $\Delta/T_K=8.6$ we find 
\begin{equation}
\begin{split}
\mathrm{d}(E_{S_z=1}-E_{S_z=0})/\mathrm{d}D &= \langle
1|S_{\mathrm{imp},z}^2|1\rangle - \langle
0|S_{\mathrm{imp},z}^2|0\rangle \\
&\approx 0.867 - 0.266 \approx 0.6,
\end{split}
\end{equation}
rather than the free-impurity result $1$. The anisotropy effects are
thus significantly renormalized by the exchange coupling of the
impurity with the host. This observation is important for the
interpretation of possible experimental results: from the
sub-gap excitation spectra, one cannot directly obtain the
``bare'' anisotropy parameters that appear in the Hamiltonian,
only their effective ``renormalized'' values.

The transitions between the ground state and the excited states
correspond to discrete (delta) sub-gap peaks in the differential
conductance spectra. We plot the spectra of these peaks in the right
panels of Fig.~\ref{fig2}. Only the transitions between the ground
state and the excited states with $\Delta S_z = \pm 1/2$ are
observable spectroscopically. Multiple sub-gap peaks may be observed,
for example, in the ``Kondo regime'' with $S_z=\pm 1/2$ ground state
for $D \neq 0$. A characteristic feature is that some sub-gap peaks
may disappear abruptly as a function of $D/\Delta$ when the ground
state changes. We also note that in an interacting superconducting
system, it is crucial to distinguish between the many-particle states
(ground state and excited states) and the peaks in the spectral
functions associated with transition between said states. Namely, to
each many-particle excitation with energy $E$, such that $\Delta
S_z=\pm 1/2$ with respect to the ground state, corresponds not one but
two spectral peaks in the single-particle spectral function. They are
located symmetrically at $\omega=\pm(E-E_\mathrm{GS})$; it is
possible, for example, to end up in exactly the same many-particle
state by either adding an electron ($\omega>0$ peak) or by removing it
($\omega<0$ peak). The lack of the distinction between the
many-particle states and the spectral peaks has led to some confusion
in the literature. The difference is particularly important for
interacting systems, where the many-particle states cannot always be
decomposed into products of single-particle levels (quasiparticles). A
notable example is the two-channel $S_\mathrm{imp}=1/2$ model.

\begin{figure}[htbp]\centering
\includegraphics[clip,width=7cm]{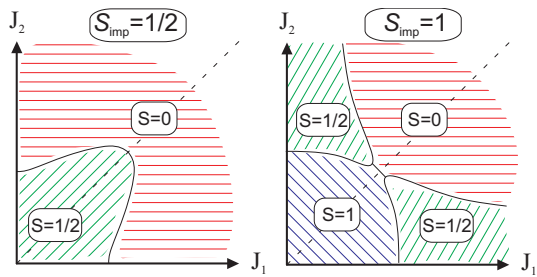}
\caption{(Color online) Schematic diagram representing the many-body
ground state as a function of the exchange coupling constants $J_1$
and $J_2$ for the isotropic ($D=0$) two-channel $S_\mathrm{imp}=1/2$
and $S_\mathrm{imp}=1$ Kondo models at fixed superconducting gap
$\Delta=10^{-5}$. Zero magnetic field, $B=0$. }
\label{phase2}
\end{figure}

In fully general multi-channel problems with non-equal $J_i$, there
are multiple stages of the Kondo screening with different Kondo
temperatures, $T_K^{(i)}$, $i=1,{\ldots},N$. Depending on the relation
between $\Delta$ and all $T_K^{(i)}$, the system may end up in
different ground states. As an illustration, in Fig.~\ref{phase2} we
depict the possible ground states for isotropic $N=2$ problems by
fixing $\Delta$ and plotting the phase diagrams in the $(J_1,J_2)$
plane. The case of $S_\mathrm{imp}=1/2$ is special due to the
overscreening effects. The second case with $S_\mathrm{imp}=1$ shows
the generic phase diagram for all $S_\mathrm{imp}\geq 1$ and $N=2$: in
the vicinity of the equal-coupling line, the ground-state spin changes
by $N/2=1$, while for general $J_i$ the multiple Kondo scales
$T_K^{(i)}$ result in intermediate regimes with only partial impurity
screening. This is a new feature that is particular to multi-channel
problems. In the presence of anisotropy $D$, the phase diagrams in
Fig.~\ref{phase2} may be extended into the third dimension; the GS
multiplets with $S\geq 1$ split according to the sign of $D$, and
effects similar to those represented in Fig.~\ref{phase1} are
observed. (See Fig.~\ref{appfig6} for some results of the calculations
at finite magnetic anisotropy $D$.)

\section{Sub-gap excited states and spectral peaks in the
multi-channel case}

\label{sec5}

\begin{figure}[htbp]\centering
\includegraphics[clip,width=7cm]{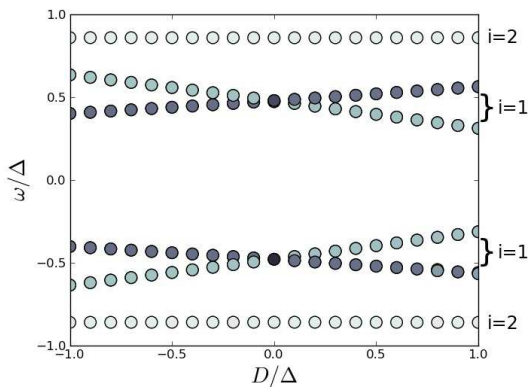}
\caption{Energies and weights of the sub-gap peaks for the
two-channel $S_\mathrm{imp}=1$ Kondo model as a function of the
magnetic anisotropy. Parameters are $\Delta=10^{-5}$, $J_1=0.2$, and
$J_2=0.1$. Darker shade corresponds to higher spectral weight. The
labels $i=1$ and $i=2$ indicate spectral peaks associated with the
$i$-th scattering channel. Zero magnetic field, $B=0$.}
\label{b2ch}
\end{figure}

For real transition-metal impurities on a surface, the exchange
coupling constants depend on the energies and the hybridization
constants of the $d$ orbitals. Assuming an adsorption site of low
symmetry, it is generally more likely that one of the exchange
constants, say $J_1$, will be dominant. In strict single-channel
problems, there is always at least one excited state inside the gap.
It tends to be located at the gap edge for $T_K \ll \Delta$ and $T_K
\gg \Delta$, but it is found well inside the gap when $T_K$ and
$\Delta$ are roughly of the same order of magnitude. In the presence
of additional scattering channels, further sub-gap excitations may
appear. For $N=2$, we find, however, that if $J_2$ is much lower than
$J_1$, the additional excited state merges with the continuum and is
not observable. For moderate $J_2/J_1$ ratio, the channel-2 excitation
peak is still inside the gap, but its spectral weight tends to be much
smaller than that in the dominant $i=1$ channel. 

On more symmetric surfaces (adsorption sites), some of the exchange
coupling constants $J_i$ can be equal, for example pair-wise equal in
the presence of a four-fold symmetry axis. In this case, the
excitations associated with those $J_i$ that are equal will be
degenerate, thus multiple sub-gap peaks are again not expected.

These results suggest that it is not very likely to observe multiple
peaks due to coupling to different scattering channels. Given that
multiple peaks are nonetheless commonly observed in experiments
\cite{ji2008, iavarone2010, ji2010}, we propose that a very likely
interpretation involves the presence of the magnetic anisotropy
splitting of the sub-gap excitations. An example for $N=2$ is shown in
Fig.~\ref{b2ch}. For $D=0$, the ground state is $S=1/2$ and there is an
$S=1$ excited state associated with channel 1 and an $S=0$ excited
state associated with channel 2. For $D\neq0$, the triplet excited
state splits. In the presence of longitudinal anisotropy, the peaks
will split at most in two, but additional splitting may be induced by
the transverse anisotropy $E(S_x^2-S_y^2)$, which is also known to be
present in adsorbed magnetic impurities \cite{hirjibehedin2007}.

\section{Behavior in the external magnetic field}
\label{sec6}

\begin{figure}[htbp]\centering
\includegraphics[clip,width=8cm]{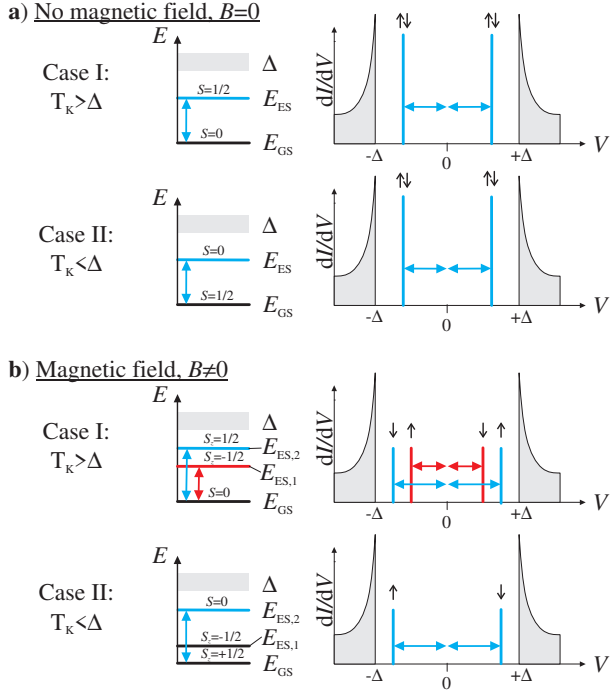}
\caption{(Color online) Schematic representation of the relation
between the many-body energy levels (shown on the left;
$E_\mathrm{GS}$ is the ground-state energy, $E_\mathrm{ES}$ are the
excited-state energies, $\Delta$ is the gap beyond which starts the
continuum of the conduction-band states) and the spectral peaks which
can be measured by the STM (shown on the right; $V>0$ indicates
tunneling into ``empty'' states, i.e., the electron addition
transitions, while $V<0$ corresponds to the removal of electrons from
the system). Only transitions where $S_z$ changes by $1/2$ are
spectroscopically visible at very low temperatures. The arrows
indicate the spin of the electron being added (positive $V$) or
removed (negative $V$) from the system at the given resonance. In the
absence of the field, $T_K>\Delta$ and $T_K<\Delta$ cases cannot be
easily distinguished spectroscopically since they both exhibit a pair
of excitation peaks at $\omega=\pm(E_\mathrm{ES}-E_\mathrm{GS})$. In
the magnetic field, we expect to observe peak splitting for $T_K >
\Delta$, while for $T_K < \Delta$ we expect merely a shift of the peak
pair concomitant with the weight reduction if the experimental
temperature is much lower than the $E_{\mathrm{ES},1}-E_\mathrm{GS}$
scale. 
The example depicted in the plot corresponds to an $S_\mathrm{imp}=1/2$
magnetic impurity in the single-channel case.} \label{magneticfield}
\end{figure}

\begin{figure}[htbp]\centering
\includegraphics[clip,width=9cm]{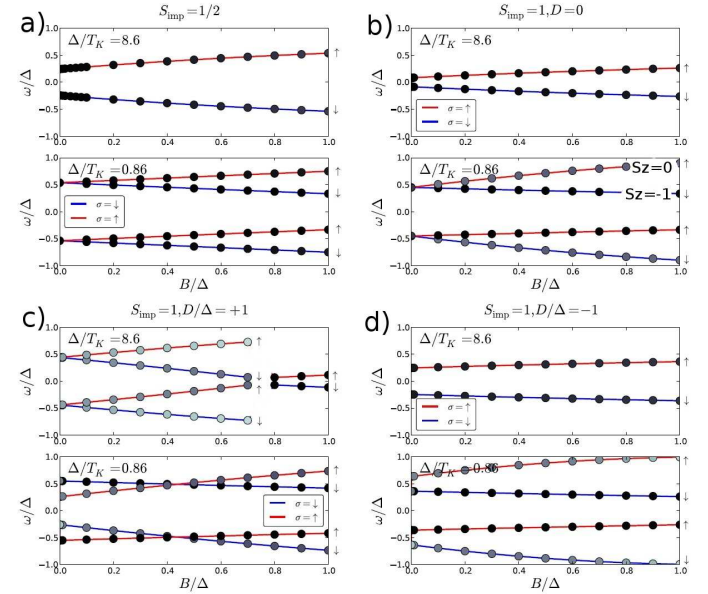}
\caption{(Color online) Energies and weights of the sub-gap spectral
peaks as a function of the external magnetic field for the
single-channel spin-$1/2$ and spin-$1$ Kondo models with different
magnetic anisotropy terms. }
\label{appfig4}
\end{figure}

Due to the strongly enhanced spin-orbit interaction on surfaces, the
interpretation of multiple peaks in terms of magnetic anisotropy
splitting appears very plausible. To experimentally distinguish
between the different possible origins of the multiple-peak sub-gap
structures in a conclusive way, we propose to study the Zeeman
splitting of the sub-gap peaks by weak magnetic fields (weak enough so
that the superconductivity is not significantly suppressed; see also
Refs.~\onlinecite{meservey1970, meservey1988, suderow2002,
rodrigo2004}). This is possible because magnetic atoms adsorbed on the
surface of a superconductor are not fully shielded by the Meissner
effect. Ultra-low-temperature STM's equipped with dilution
refrigerators are likely to achieve sufficient energy resolution. The
magnetic field splits pairs of the sub-gap states with the same
$|S_z|$. Since the only observable transitions are those with $\Delta
S_z= \pm 1/2$, each spectral peak may split at most in two. In the
presence of transverse anisotropy and/or transverse magnetic field,
the spin symmetry is fully lifted and even more complex spectra of
excitation peaks can arise. We note, however, that if the ground state
is spin degenerate (i.e., for partially screened impurity), even a
small magnetic field can fully polarize the residual impurity spin,
thus some of the sub-gap peaks might not be spectroscopically
observable since their weights are essentially zero (at $T=0$). Some
observed peaks will thus merely shift, rather than split. This
behavior is represented schematically in Fig.~\ref{magneticfield} for
the example of an $S_\mathrm{imp}=1/2$ impurity in the single-channel
case, while the results of a corresponding NRG calculation are shown
in Fig.~\ref{appfig4}(a).

The single-channel $S_\mathrm{imp}=1$ model is studied in
Fig.~\ref{appfig4}b,c,d) for different values of the magnetic
anisotropy $D$. In the absence of the anisotropy, $D=0$, the behavior
is similar to that in the spin-$1/2$ case: for $\Delta/T_K > 0$
(``free-spin'' regime), the peaks merely shift, while for
$\Delta/T_K<0$ (Kondo regime), there is a splitting of the peaks, since
the transitions are possible from the low-lying doublet state (here
$S_z=-1/2$) to both $S_z=-1$ and $S_z=0$ excited states. It is worth
emphasizing that in the latter case, the peak weights are different for
the two transitions. A simplified, but intuitive picture is the
following: the ground state is a Kondo state
$\ket{S_{\mathrm{imp},z}=-1,\uparrow}$, that is, a bound state of an
$S_z=-1$ impurity state and a spin-up conduction-band electron with
total $S_z=-1/2$. The first excited state has $S_z=-1$ and can be
obtained in the process of adding a spin-down electron. The second
excited state has $S_z=0$ and it can be reached by adding a spin-up
electron. Since there is already a spin-up electron present in the
Kondo ground state, adding a further spin-up electron will have a
reduced weight as compared to adding a spin-down electron. This
behavior persists in the presence of magnetic anisotropy: the peak
corresponding to a transition to the $S_z=0$ state has lower weight
than the peak corresponding to the transition to the $S_z=-1$ state in
all three cases of $D=0$, $D>0$, and $D<0$; see
Fig.~\ref{appfig4}b,c,d).

In the ``free-spin'' regime, the evolution of the spectral peaks in
the magnetic field depends on the sign of the magnetic anisotropy;
compare the upper panels in Fig.~\ref{appfig4}b,c,d). The isotropic
$D=0$ and easy-axis $D<0$ cases are similar: in the presence of the
field, the $S_z=-1$ state will be the ground state and the transitions
are only possible to the $S_z=-1/2$ excited state, thus a single
sub-gap peak pair is observed. The easy-plane $D>0$ case is more
interesting. Now the ground state (for small magnetic fields) is
$S_z=0$, thus the transitions to both $S_z=+1/2$ and $S_z=-1/2$
excited states are possible; we thus see two pairs of sub-gap spectral
peaks, that is, a total of four sharp peaks. For very large Zeeman
splitting, the $S_z=-1/2$ excited state will become the new ground
state of the system; at this point only the transition to the $S_z=0$
sub-gap will be possible and a single pair of peaks will remain in the
impurity spectral function.

\begin{figure}[htbp]\centering
\includegraphics[clip,width=8.5cm]{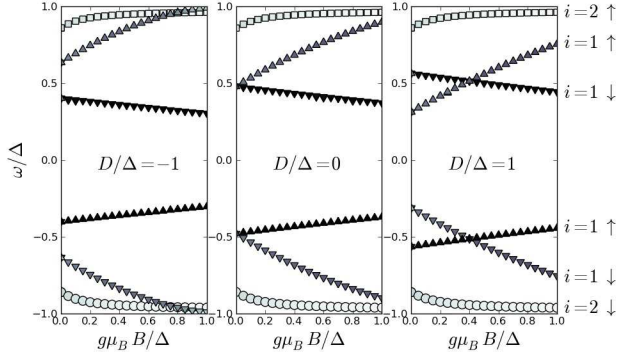}
\caption{Energies and weights of the sub-gap spectral peaks as a
function of the magnetic field $B$ for the two-channel $S_\mathrm{imp}=1$
impurity. Parameters are as in Fig.~\ref{b2ch}. The spectral peaks are
fully spin-polarized, as indicated by the arrows.} \label{fig4}
\end{figure}

In Fig.~\ref{fig4}, we finally plot the field dependence of the
spectral peaks of a two-channel spin-1 impurity for different magnetic
anisotropies $D$. In all three cases, the ground state for $B > 0$ is
non-degenerate, $S_z=-1/2$, and one may only observe the transitions
to $S_z=-1$ and $S_z=0$ excited states. As evidenced by the results in
Fig.~\ref{fig4}, the sign and magnitude of $D$ can be easily
determined from the shifts. The peak due to the second weakly coupled
conduction channel is always weaker as compared to the peaks
associated with the dominant screening channel.

\section{Effects of the inter-impurity exchange interaction}
\label{sec7}

When adsorbed magnetic impurities are brought together, for example by
controlled manipulation using the tip of the STM, their mutual
interaction will change the LDOS signatures measured by the scanning
tunneling spectroscopy \cite{chen1999, madhavan2002, lee2004,
  hirjibehedin2006, wahl2007, ji2008}. The theory of the
inter-impurity interactions in the normal case has been worked out
using both simplified model Hamiltonians and ab initio
calculations \cite{lau1978, stepanyuk1996, hyldgaard2000,
  stepanyuk2003co, strandberg2007}. In the superconducting case, the
calculations of the sub-gap excitation spectra in impurity dimers have
been performed mainly for classical impurity spins \cite{flatte2000,
  morr2003}.  Recently, some calculations for coupled quantum
impurities have been performed in the context of double quantum dots
\cite{dqdsc}. Here we study the problem of two quantum impurities,
each coupled to a separate conduction band with a superconducting gap
(with equal $\Delta$), and interacting via an isotropic Heisenberg
Hamiltonian:
\begin{equation}
H_\mathrm{int} = J \vc{S}_{\mathrm{imp},1} \cdot
\vc{S}_{\mathrm{imp},2}.
\end{equation}

\begin{figure}[htbp]\centering
\includegraphics[clip,width=8cm]{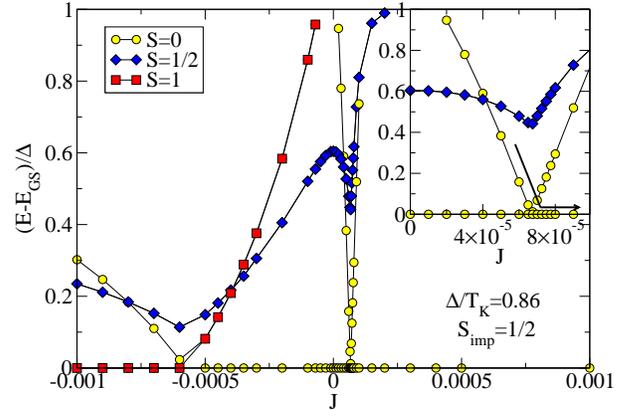}
\caption{(Color online) Ground state and sub-gap many-body excited
states in the two-impurity problem with $S_\mathrm{imp}=1/2$
impurities coupled by Heisenberg exchange coupling $J$. The inset is a
close-up view of the region where the two different singlet ground
states cross: the arrow shows the evolution of the local AFM-ordered
spin-singlet ground state as $J$ is increased. The gap is
$\Delta=10^{-5}$.}
\label{figint1}
\end{figure}

We first discuss the simplest case of two $S_\mathrm{imp}=1/2$
impurities. The numerical results for the sub-gap states are shown in
Fig.~\ref{figint1}. The parameters are chosen such that in the absence
of coupling, each impurity is the Kondo screened regime with an $S=0$
ground state and one $S=1/2$ sub-gap excited state (two-fold
degenerate due to spin); for the two-impurity system, the ground state
is thus a singlet (we will call it the ``Kondo singlet''), and there
are two degenerate $S=1/2$ excited states (four-fold total
degeneracy). As the antiferromagnetic ($J>0$) exchange coupling is
turned on, a new singlet state emerges from the continuum; this state
can be interpreted as arising from the antiferromagnetically ordered
local singlet state formed by the two impurity spins, and we will
therefore denote it as the ``AFM singlet''. It should be emphasized
that this state is not directly related to the $S=1/2$ excited states
(which are actually present in the sub-gap spectrum at the same time
as the new AFM singlet); the AFM singlet state should thus not be
interpreted as emerging from the coupling of the $S=1/2$ sub-gap
localized states, but rather as arising from the local inter-impurity
singlet state.  We furthermore emphasize that the two $S=1/2$ excited
states remain degenerate (this holds, in fact, for all values of $J$).
As $J$ is increased beyond $J \approx 6 \times 10^{-5}$, the Kondo and
AFM singlet states cross and exchange their roles as the ground and
the excited state. At still higher $J$, both the excited Kondo singlet
state and the two $S=1/2$ excited states merge with the continuum and
are no longer observable. For large antiferromagnetic inter-impurity
coupling, the dimer behaves as a non-magnetic object and therefore
does not have any excitations deep inside the gap. Such behavior seems
to be present for Cr dimers (in the configuration ``Cr dimer II'' with
small inter-atom separation) \cite{ji2008}. 

Equally interesting is the case of ferromagnetic Heisenberg coupling,
also shown in Fig.~\ref{figint1}. In this case, a new ``FM triplet''
sub-gap state emerges in the sub-gap spectrum. This state decreases in
energy until it replaces the Kondo singlet as the new ground state.
Two important observation can be made: (i) the level crossing between
the FM triplet and the Kondo singlet occurs for a much larger (by an
order of magnitude) absolute value of the Heisenberg coupling as the
level crossing between the AFM singlet and the Kondo singlet; (ii) with
increasing $|J|$, the Kondo singlet and the degenerate $S=1/2$ excited
states evolve only slowly and remain deep inside the gap even for very
large ferromagnetic exchange coupling. This suggests that in the case
of ferromagnetic dimers, we are more likely to observe some sub-gap
spectral peaks. It is possible that the Mn dimers (even in the
configuration ``Mn dimer II'' with small inter-atom separation
\cite{ji2008}) have ferromagnetic spin coupling.

\begin{figure}[htbp]\centering
\includegraphics[clip,width=8cm]{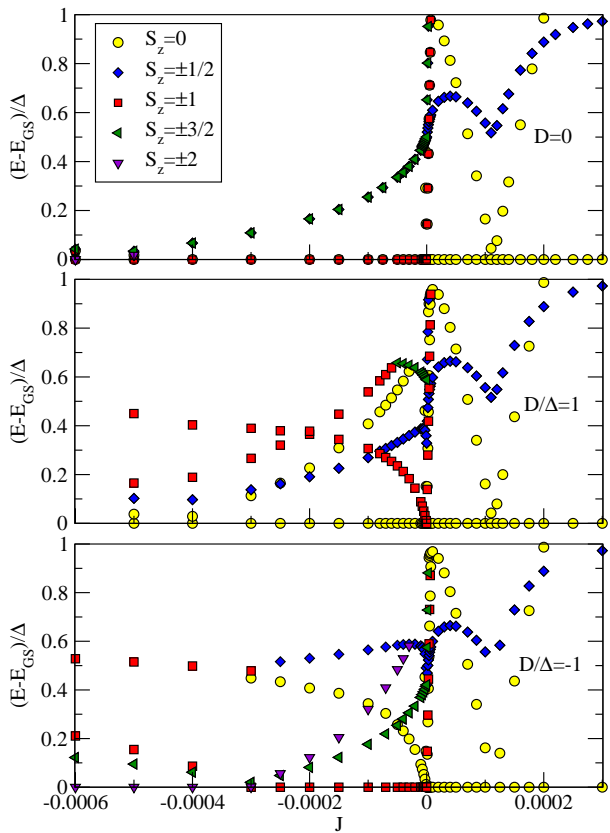}
\caption{(Color online) Ground state and sub-gap many-body excited
states in the two-impurity problem with $S_\mathrm{imp}=1$
impurities coupled by Heisenberg exchange coupling $J$.
The gap is $\Delta=10^{-5}$.}
\label{figint2}
\end{figure}

\begin{figure}[htbp]\centering
\includegraphics[clip,width=8cm]{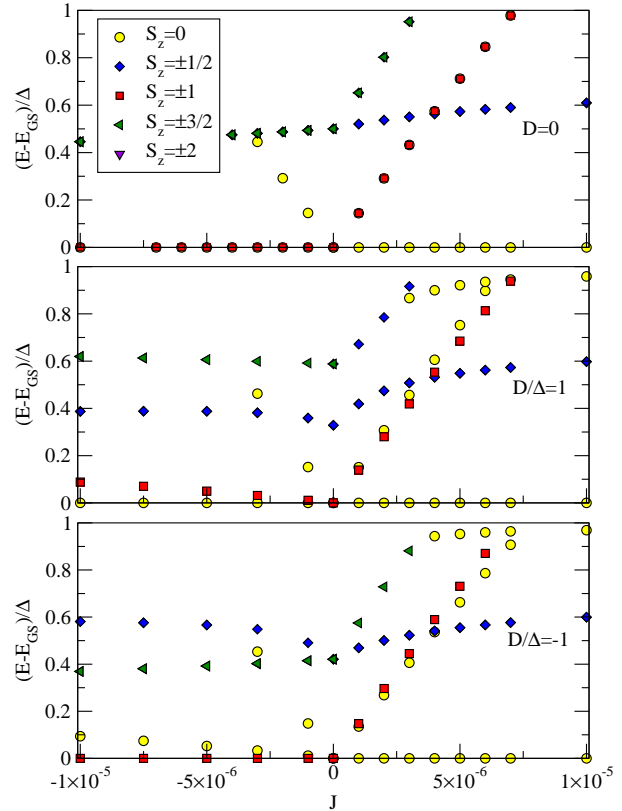}
\caption{(Color online) Close-up view of the small-$J$ region in
Fig.~\ref{figint2}.}
\label{figint2bis}
\end{figure}

We now turn to the case of two isotropic $S_\mathrm{imp}=1$
impurities, each coupled to a single channel; see the upper panel in
Fig.~\ref{figint2}. We again consider the parameter regime where each
impurity is Kondo-screened. This time, however, the impurities have
residual spin $1/2$. In the absence of the inter-impurity coupling,
the ground state is degenerate and composed of a singlet and a triplet
``Kondo state'', and there are degenerate excited states with $S=1/2$
and $S=3/2$. The ground-state degeneracy is lifted by a small $J$,
thus for small $J$ the ground state is either a singlet (for $J>0$) or
a triplet (for $J<0$); this splitting occurs on the energy scale of
almost ``bare'' $J$; see the close-up view in Fig.~\ref{figint2bis}. For
antiferromagnetic exchange coupling, there is an additional singlet
state that arises from the local inter-impurity singlet state in
which two $S=1$ spins are rigidly antiferromagnetically ordered. This
state is different from the singlet Kondo state, which arises from AFM
ordering between two residual $S=1/2$ extended objects. This is
another confirmation of the nature of the sub-gap singlet states, which
we had already discussed in the case of $S_\mathrm{imp}=1/2$. For
sufficiently large AFM coupling, the singlet states cross, and for very
large coupling we again find that there are no sub-gap excited states.
Note also the similarity between the $J>0$ behavior in
$S_\mathrm{imp}=1/2$ and $S_\mathrm{imp}=1$ models.

For ferromagnetic coupling, the situation is again similar to what we
had observed for $S_\mathrm{imp}=1/2$ impurities: the
ferromagnetically ordered $S=2$ local object becomes the new ground
state only for very large Heisenberg coupling. The transition occurs
at $J \approx -6\times 10^{-5}$, that is, at essentially the same value
as for $S_\mathrm{imp}=1/2$. In addition, we again observe that some
sub-gap excited states are typically found in the sub-gap spectrum
even for large $|J|$.

Finally, we discuss how the sub-gap excitation spectra change in the
presence of the magnetic anisotropy. In the two lower panels in
Fig.~\ref{figint2}, we show the results for easy-plane and easy-axis
magnetic anisotropy for the $S_\mathrm{imp}=1$ impurities. The results
are in general similar to those for the isotropic model, but there are
some notable differences. 

For AFM coupling , we find that the ground state is an $S_z=0$ state
(i.e. , the ``singlet'' state) and that there are no sub-gap excited
states for large enough $J$. In fact, it may be noticed that the
results for $J>0$ are remarkably similar irrespective of the sign and
strength of the magnetic anisotropy.

For small ferromagnetic exchange coupling, the $S=1$ ground state
splits: for easy-plane anisotropy the ground state is $S_z=0$, while
for easy-axis anisotropy it is $S_z=\pm1$. Thus the combined
inter-impurity states can be understood within a two-stage splitting
model: in the first step, we consider how the states combine due to the
Heisenberg exchange coupling; in the second step, we consider how
the resulting states split due to the magnetic anisotropy. Such
separation is possible because in the parameter regime under
consideration, the scale of the exchange splitting (up to several times
$10^{-4}$) is larger than the scale of the magnetic anisotropy (fixed
at $|D|=10^{-5}$). For strong FM coupling, the result depends on the
type of anisotropy: the ground state arises from an $S=2$ state, but
for the easy-plane case the actual ground state is a non-degenerate
$S_z=0$ state, while for the easy-axis case the ground state is a
twofold-degenerate $S_z=\pm 2$ state. The transition between the
small-$|J|$ and large-$|J|$ ground state occurs at a value of $J$
that depends on the magnetic anisotropy.

\section{Conclusion}

We have studied single magnetic impurities on superconductor surfaces
as well as their dimers. We have emphasized the importance of using
quantum impurity models to describe the sub-gap excitation spectrum of
the magnetic adatoms. The isotropic models exhibit a transition between
ground states with different degrees of Kondo screening of the
impurity spin; this depends on the values of the Kondo exchange
coupling constants and the ratios between the resulting Kondo
temperatures and the BCS superconducting gap. The sub-gap excited
states only appear in cases where at least one of the Kondo
temperature scales is of the same order as the superconducting gap. If
all coupling constants are small (i.e., if the Kondo effect is
absent), there are no sub-gap excited states deep inside the gap, which
is at odds with the predictions from the classical spin models. The
ground state and sub-gap excited states with $S \geq 1$ are split in
the presence of the magnetic anisotropy. It is found that the
splitting is strongly renormalized by the Kondo screening, thus the
model needs to be studied using non-perturbative techniques, such as
the numerical renormalization group. We find that the weight of the
sub-gap peaks is the largest for strongly coupled Kondo channels, but
becomes lower for weakly coupled channels. In addition, the sub-gap
excitations associated with weakly coupled channels tend to appear
near the gap edge, and may therefore be difficult to observe. This
finding suggests that multiple-peak sub-gap excitations likely arise
from the internal structure of the impurity (spin-orbit coupling
leading to the magnetic anisotropy). We have also explored the
inter-dot exchange coupling for impurity dimers: for strong
antiferromagnetic coupling, no sub-gap excitations are present, while
even for relatively strong ferromagnetic coupling some sub-gap peaks
may be observable. We have shown that the external magnetic field has
a sizable effect on the sub-gap excitation spectra of impurities.
Exploring the magnetic properties of impurities using the
field dependence of the sub-gap peaks constitutes a worthwhile
experimental challenge. Furthermore, using a break-junction setup with
superconducting contacts should make it possible to map the excitation
spectra in the entire $(D,B)$ plane at the same time, providing
further means to test the predictions of this work.

\begin{acknowledgments}
RZ would like to thank E. Tosatti and M. Fabrizio for interesting
discussions. R.Z. acknowledges the support of the Slovenian Research
Agency (ARRS) under Grant No. Z1-2058. T.P. acknowledges support by the
Deutsche Forschungsgemeinschaft through SFB 602.
\end{acknowledgments}

\appendix

\section{Sub-gap excitation spectra}

\begin{figure*}[htbp]
\centering
\includegraphics[clip,width=15cm]{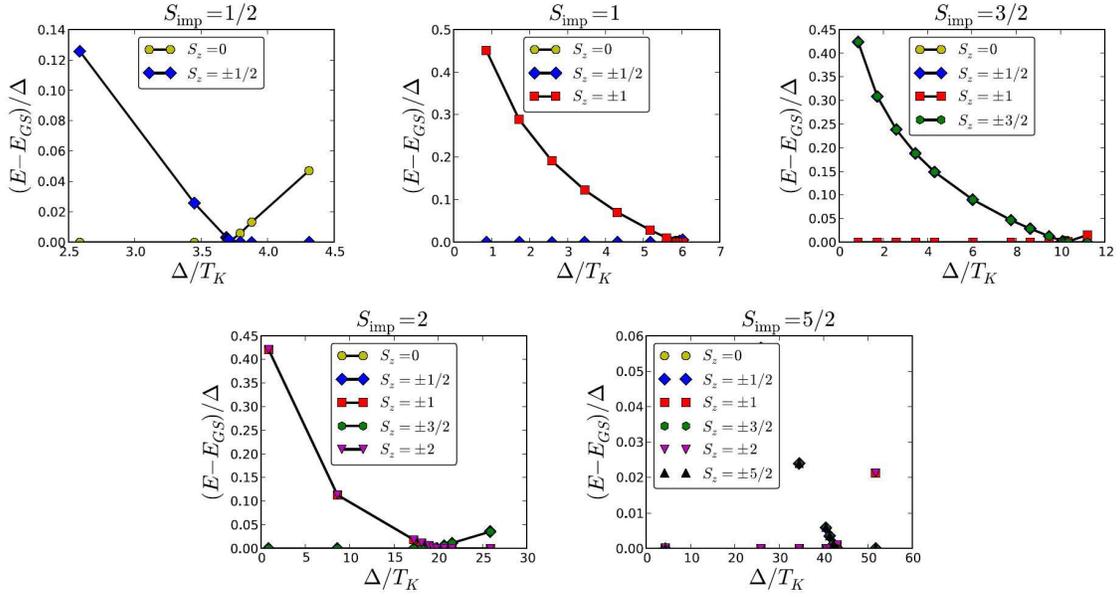}
\caption{(Color online) Ground state and sub-gap many-body excited states as a
function of the superconducting gap $\Delta$ for single-channel
($N=1$) Kondo models with different impurity spin $S_\mathrm{imp}$.
The NRG calculations have been performed with the NRG discretization
parameter $\Lambda=2$. Additional $\Lambda$-scaling calculations
suggest that the results for the transition point $\Delta_c/T_K$ given
in Table~1 are accurate within few percent. } \label{appfig1}
\end{figure*}

\begin{figure*}[htbp]
\centering
\includegraphics[clip,width=15cm]{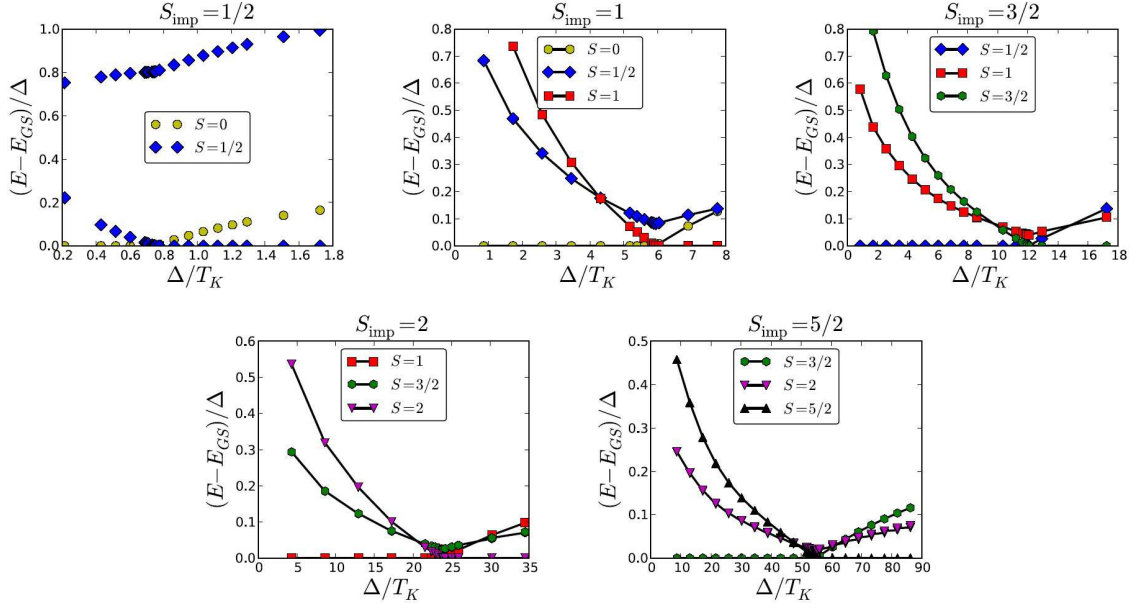}
\caption{(Color online) Ground-state and sub-gap many-body excited
states as a function of the superconducting gap $\Delta$ for
two-channel ($N=2$) Kondo models with different impurity spin
$S_\mathrm{imp}$.  
The NRG calculations have been performed with the discretization
parameter $\Lambda=4$. The results given in Table~1 for the
two-channel case have been calculated by performing further
calculations using a twist parameter $z=0.5$ and averaging the results
for $z=1$ and $z=0.5$. The tabulated results are then accurate within
a few percent for both the single-channel and the two-channel case. Note the
presence of the additional $S=S_\mathrm{imp}-1/2$ excited-state
multiplet inside the gap. For non-equal $J_i$, this excited state can
become the ground state in some parameter regimes; see
Fig.~\ref{phase2}.} \label{appfig2}
\end{figure*}

\begin{figure*}[htbp]\centering
\includegraphics[clip,width=15cm]{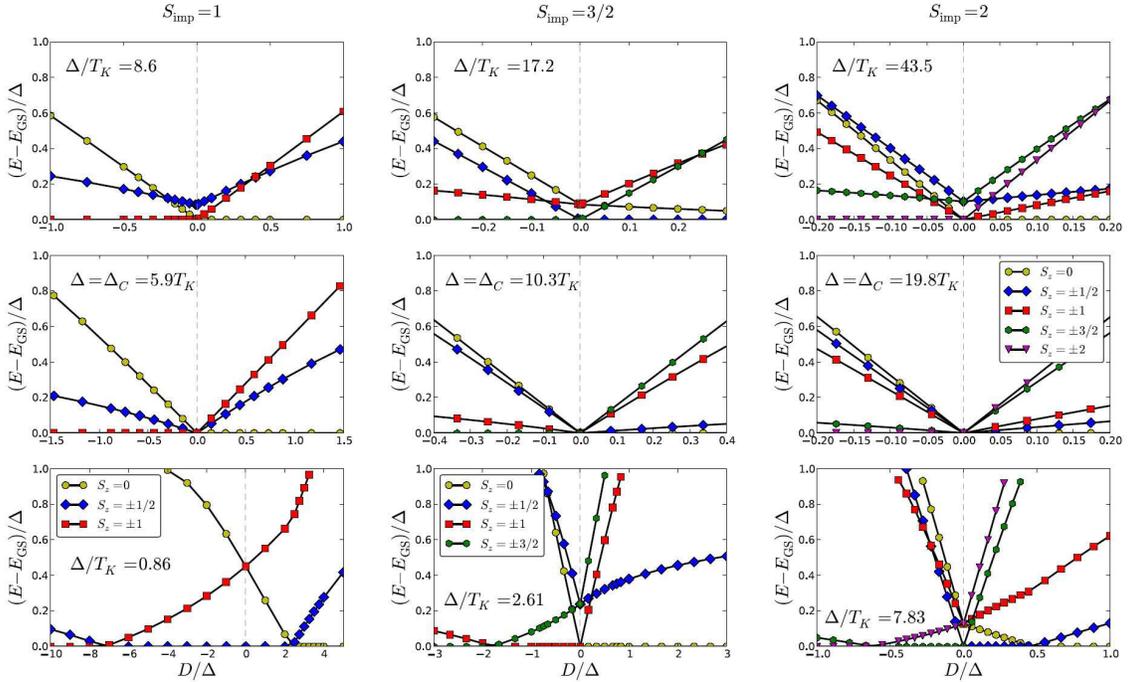}
\caption{(Color online) Ground-state and sub-gap excited states as a
function of the magnetic anisotropy $D$ for different values of the
superconducting gap $\Delta$ for the single-channel spin-$1$,
spin-$3/2$, and spin-$2$ Kondo models. The results of these (and
similar) calculations have been used to establish the schematic phase
diagrams shown in Fig.~\ref{phase1}.}
\label{appfig3}
\end{figure*}

\begin{figure*}[htbp]\centering
\includegraphics[clip,width=15cm]{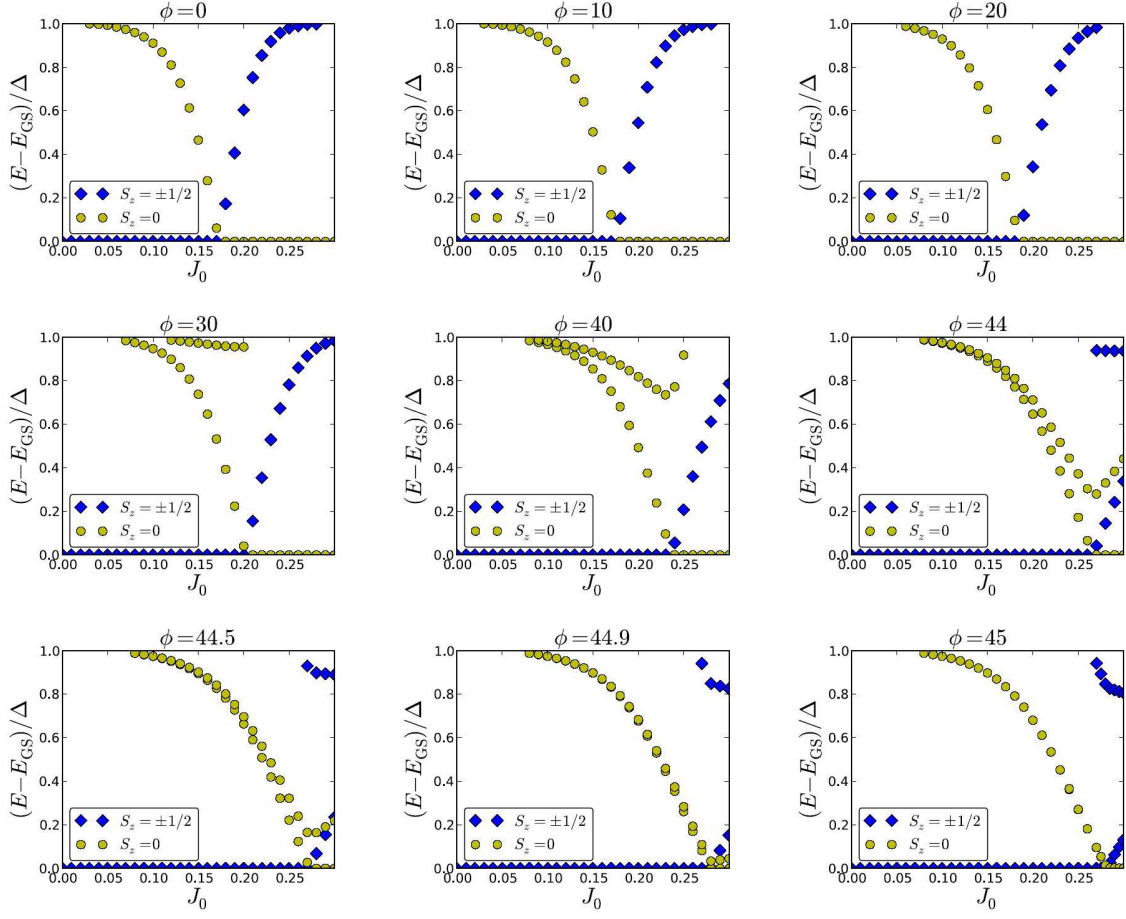}
\caption{(Color online) Ground state and sub-gap excitations as a function
of the parameter $J_0$ for the two-channel $S_\mathrm{imp}=1/2$ Kondo
model. We define $J_1=J_0 \cos\phi$ and $J_2=J_0 \sin\phi$; the angles
$\phi$ of the direction in the $(J_1,J_2)$ plane are given (in
degrees) as the titles of the subfigures. The gap is $\Delta=10^{-5}$.
The results of these (and similar) calculations have been used to
establish the schematic phase diagram for the $S_\mathrm{imp}=1/2$
case in Fig.~\ref{phase2} of the main text. The two-channel
calculations with non-equal $J_i$ have to be performed with a higher
value of the NRG discretization parameter $\Lambda=8$, otherwise they
are intractable. No twist-parameter averaging has been performed here;
nevertheless, the results are still qualitatively correct.}
\label{appfig5}
\end{figure*}

\begin{figure*}[htbp]\centering
\includegraphics[clip,width=10cm]{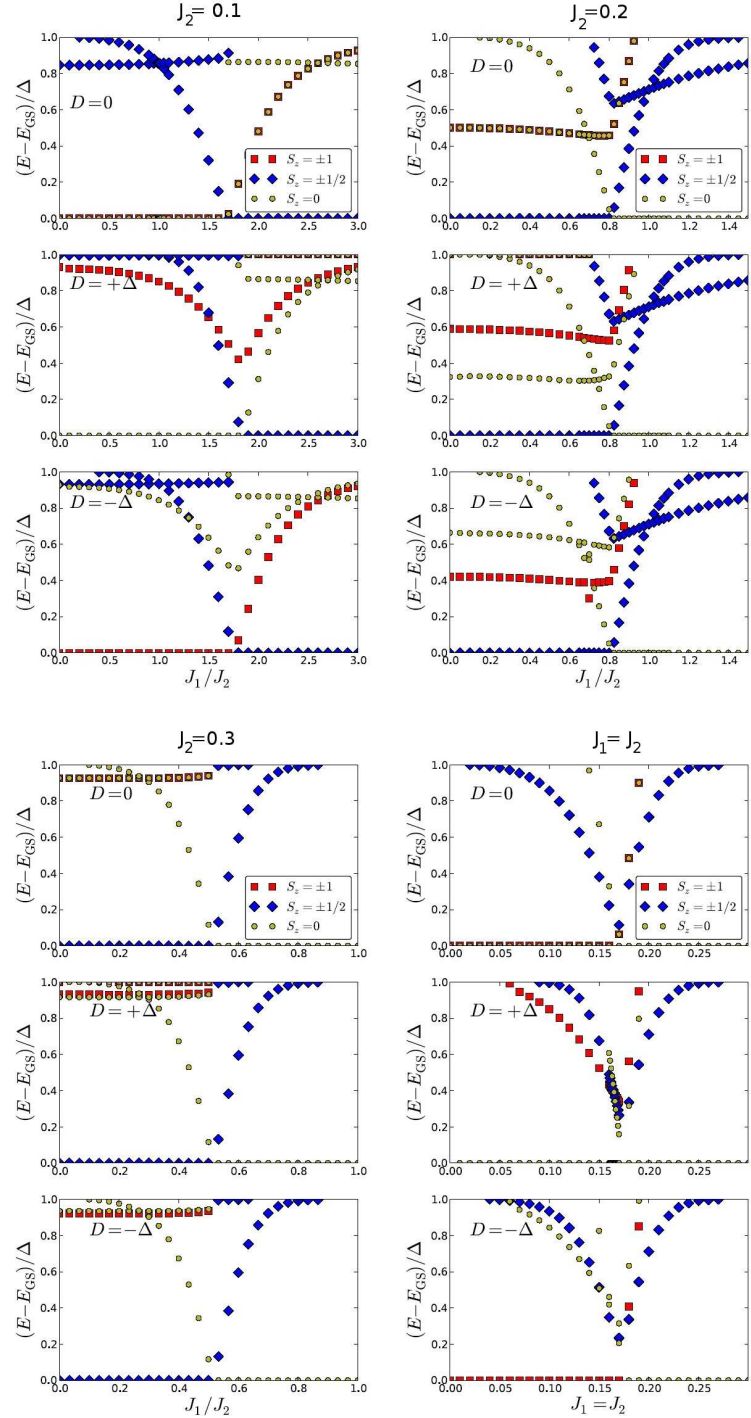}
\caption{(Color online) Ground state and sub-gap many-particle
excitations as a function of the parameters $J_1$ and $J_2$ for the
two-channel $S_\mathrm{imp}=1$ Kondo model. The gap is
$\Delta=10^{-5}$.
The results of these (and similar) calculations have been used to
establish the schematic phase diagram for the $S_\mathrm{imp}=1$ case
in Fig.~\ref{phase2} of the main text. }
\label{appfig6}
\end{figure*}

This appendix contains additional figures detailing the results
discussed in the main text. Figures~\ref{appfig1} and \ref{appfig2}
show the sub-gap states for one-channel and two-channel Kondo models
from which the transition points between the (partially) screened and
unscreened regimes can be extracted; see Table~\ref{tab1} and
Figure~\ref{phase1} in the main text. Fig.~\ref{appfig3} shows the
effect of the magnetic anisotropy on the sub-gap states, that is, the
splitting of the degenerate spin multiplets. Figures~\ref{appfig5} and
\ref{appfig6} contain the results for the two-channel $S=1/2$ and
$S=1$ Kondo model in the ($J_1$, $J_2$) plane, which serve to
establish the phase diagram in Fig.~\ref{phase2}.

\bibliography{aniso-sc}

\end{document}